\documentclass[aps,prl,preprint,groupedaddress,showpacs,amsmath,amssymb]{revtex4}
\usepackage{graphicx}% Include figure files
\usepackage{dcolumn}% Align table columns on decimal point
\usepackage{bm}% bold math

\begin{document}

\title{New Phases of $SU(3)$ and $SU(4)$ at Finite Temperature}

\author{Joyce C. Myers}
\email{jcmyers@wustl.edu}
\author{Michael C. Ogilvie}
\email{mco@wuphys.wustl.edu}
\affiliation{%
Department of Physics, Washington University, St.\ Louis, MO 63130, USA
}%

\date{\today}% It is always \today, today,

%\maketitle
\begin{abstract}
The addition of an adjoint Polyakov loop term to the action of a pure
gauge theory at finite temperature leads to new phases of $SU(N)$
gauge theories. For $SU(3)$, a new phase is found which breaks $Z(3)$
symmetry in a novel way; for $SU(4)$, the new phase exhibits spontaneous
symmetry breaking of $Z(4)$ to $Z(2)$, representing a partially
confined phase in which quarks are confined, but diquarks are not.
The overall phase structure and thermodynamics is consistent with
a theoretical model of the effective potential for the Polyakov loop
based on perturbation theory.
\end{abstract}


\maketitle

\section{Introduction}

It is well established that $SU(N)$ gauge theories in $3+1$ dimensions
have a low-temperature phase in which quarks are confined, and a high
temperature phase where quarks are deconfined, often referred to
as the quark-gluon plasma phase. The deconfinement phase transition
in pure gauge theories, {\it i.e.}, without quarks,
is understood theoretically as a transition between a low-temerature
phase where a global $Z(N)$ symmetry is unbroken to a high-temperature
phase where $Z(N)$ symmetry is spontaneously broken \cite{Svetitsky:1982gs}.
Simulations indicate that the transition from confined phase to deconfined
phase is similar for all $N \geq 3$. The global $Z(N)$ symmetry appears
to always break completely, with no residual unbroken subgroup.

The addition of a term of the form\begin{equation}
-\int d^{3}x\, h_{A}\, Tr_{A}P(\vec{x})=-T\int_{0}^{\beta}dt\int d^{3}x\, h_{A}\, Tr_{A}P(\vec{x})\end{equation}
to the Euclidean action of pure $SU(N)$ gauge theories at finite
temperature leads to new phases with novel properties. Here $P(\vec{x})$
is the Polyakov loop at the spatial point $\vec{x}$, given by the
usual path-ordered exponential of the temporal component of the gauge field
 $A_{0}$ in the Euclidean time direction.
The temporal origin of $P$ is irrelevant due to the trace; because
the trace is in the adjoint representaton, this additional term respects
$Z(N)$ symmetry. 
Of course, this additional term in the action is
neither local nor renormalizable in $3+1$ dimensions. 
Thus we must regard this model as an effective theory
defined at fixed lattice spacing
or by some other cut-off. There will be a finite renormalization of the
parameter $h_A$ in comparing lattice results with continuum.

This additional term directly changes the effective potential.
For a pure $SU(N)$ gauge theory, the effective potential 
$V_{eff}$ can be written
as a character expansion of the form
\begin{equation}
V_{eff}=\sum v_{R}Tr_{R}P\end{equation}
where the sum is over all representations of zero $N$-ality, i.e.,
invariant under $Z(N)$. 
Terms of this form can be induced at one loop
by certain topological
excitations \cite{Davies:1999uw,Davies:2000nw,Diakonov:2004jn}
as well as by
particles in the adjoint representation.
A one-loop calculation shows that the contribution
to the effective potential of a heavy particle of mass $M$ in the
adjoint representation, either boson or fermion can be approximated
 in $3+1$ dimensions as
\begin{equation}
-\left[\frac{(2s+1)M^{2}T^{2}}{\pi^{2}}K_{2}\left(M/T\right)\right]Tr_{A}\left(P\right)=-Th_{A}Tr_{A}\left(P\right)
\end{equation}
where $T$ is the temperature and $2s+1$ accounts for spin  \cite{Meisinger:2001fi}.
The parameter $h_{A}$ is positive in this case. The effect of such particles
can be included at lowest order in $h_A$
in the effective potential by the shift $v_{A}\rightarrow v_{A}-Th_{A}$.
A positive value of $h_{A}$ favors the $Z(N)$-breaking deconfined
phase. However, a term with $h_{A}$ negative favors minimization of
$Tr_{A}P$. Because $Tr_{A}P=\left|Tr_{F}P\right|^{2}-1$, the
minimization $Tr_{A}P$ of  implies
$Tr_{F}P=0$, a definining property of the confined phase. It is reasonable
to expect that a sufficiently negative value of $h_{A}$ might lead
to a restoration of confinement at temperatures above the deconfinement
temperature. 

We were motivated to look for this symmetry restoration by recent
theoretical work on various aspects of the Polyakov loop effective
potential. In certain supersymmetric gauge theories on $R^{3}\times S^{1}$,
Davies {\it et al.}  \cite{Davies:1999uw,Davies:2000nw} have shown that finite temperature monopoles give
rise to a Polyakov loop effective potential that has a $Z(N)$-symmetric
minimum for all values of the $S^{1}$ circumference, and is therefore
in a confined phase. These models do not precisely represent systems
at finite temperature, because the supersymmetric partners of the
gauge fields obey periodic boundary conditions. Comparable calculations
in non-supersymmetric $SU(N)$ gauge theories at finite temperature
are much more difficult. In $SU(2)$ gauge theory, Diakonov {\it et al.}
  \cite{Diakonov:2004jn} have calculated the contribution to $V_{eff}$ of finite-temperature
instantons with non-trivial holonomy; such instantons have a color
magnetic monopole content. Their work indicates an instability of
the deconfined phase at sufficiently low temperature. In both of these
examples, topological excitations give rise to a term in the effective
potential corresponding to $h_{A}$ negative.

A positive value of $h_{A}$ decreases the deconfinement
temperature. For negative values of $h_{A}$, we have found new phases
for both $SU(3)$ and $SU(4)$. In the case of $SU(3)$, the new phase
breaks $Z(3)$ symmetry in an unfamiliar way, 
characterized by a negative value for the Polyakov
loop in the fundamental representation $\left\langle Tr_{F}P\right\rangle <0$.
In the case of $SU(4)$, the global $Z(4)$symmetry is spontaneously
broken to $Z(2)$. The residual $Z(2)$ symmetry ensures
that for the fundamental representation $\left\langle Tr_{F}P\right\rangle =0$,
but $\left\langle Tr_{R}P\right\rangle \neq0$ for representations $R$
 that transform trivially under $Z(2)$, such as the ${\bf 6}$ and the ${\bf 10}$.

\section{Simulation Results for $SU(3)$}

\begin{figure}
\centering
\includegraphics[width=.55\textwidth]{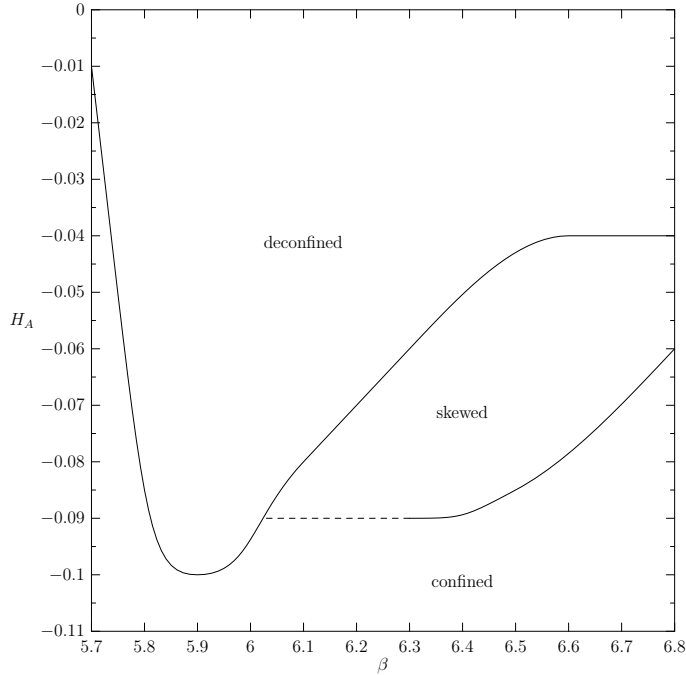}
\caption{$SU(3)$ phase diagram in the $\beta-H_A$ plane. The dotted line represents a plausible extrapolation.}
\label{fig:1}
\end{figure}

The lattice action we have studied for $SU(3)$ and $SU(4)$ is
\begin{equation}
S=S_{W}+\sum_{\vec{x}}\, H_{A}\, Tr_{A}P(\vec{x})\label{action}
\end{equation}
where $S_{W}$ is the Wilson action, defined conventionally
as the sum over plaquettes. 
The sum in the second term is over all spatial sites,
and naively $H_{A}=h_{A}a^{3}$. 
Most of our simulations were performed on
$24^{3}\times4$ lattices as reported here, but similar results were
obtained for $N_{t}=2$ and $6$. The programs used for these simulation
were developed using the programming framework FermiQCD \cite{DiPierro:2005qx}. Because
the augmented lattice action $S$ depends quadratically on the time-like
link variable $U_{0}$ via the adjoint representation, the efficient
heatbath methods developed for the standard lattice action cannot
be used. We have used instead a recently developed $SU(N)$ overrelaxation
algorithm \cite{deForcrand:2005xr} combined with the Metropolis algorithm. 
The overrelaxation algorithm,
which operates on the full $SU(N)$ group rather than subgroups, proved
to be fast and effective. Other algorithms which have been developed for
fundamental plus adjoint actions could also be used 
\cite{Hasenbusch:2004yq,Bazavov:2005vr}.
A typical simulation on a $24^{3}\times4$ lattice
consisted of 10,000 equilibration sweeps followed by 60,000 sweeps
during which 2000 measurements were performed.

\begin{figure}
\centering
\includegraphics[width=.35\textwidth]{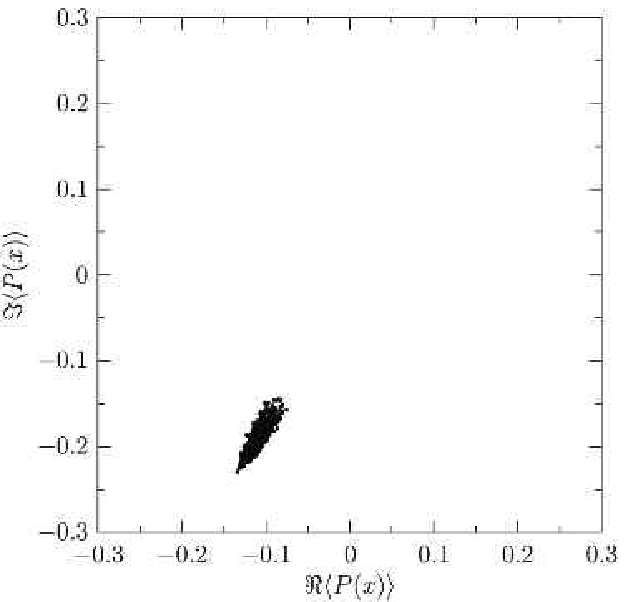}
\vspace{-5mm}
\caption{$SU(3)$ Polyakov loop histogram at $\beta=6.5$, $H_A=-0.05$.}
\label{fig:2}
%\end{figure}
\vspace{2mm}
%\begin{figure}
\centering
\includegraphics[width=.35\textwidth]{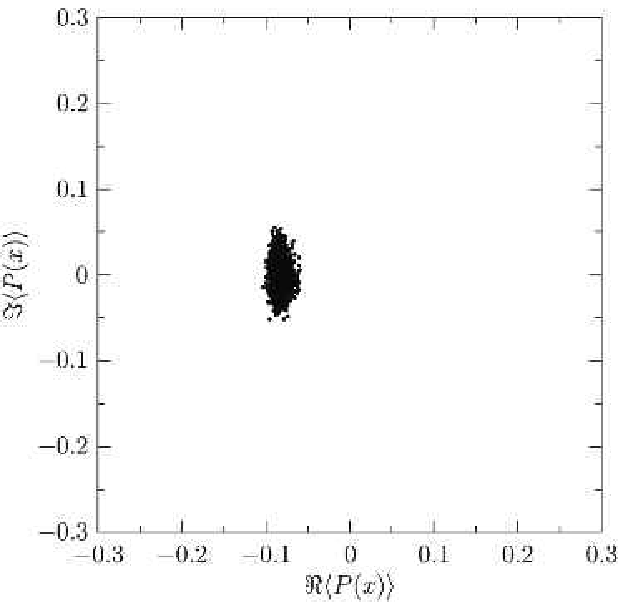}
\vspace{-5mm}
\caption{$SU(3)$ Polyakov loop histogram at $\beta=6.5$, $H_A=-0.06$.}
\label{fig:3}
%\end{figure}
\vspace{2mm}
%\begin{figure}
\centering
\includegraphics[width=.35\textwidth]{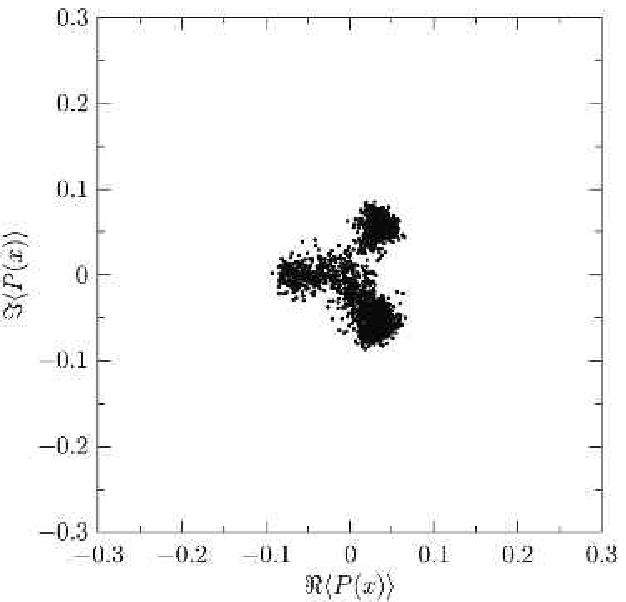}
\vspace{-5mm}
\caption{$SU(3)$ Polyakov loop histogram at $\beta=6.5$, $H_A=-0.08$.}
\label{fig:4}
\end{figure}

The approximate phase diagram for $SU(3)$ is shown in Fig. \ref{fig:1} for
$N_{t}=4$. The order parameter is $Tr_{F}P$,
projected for each  lattice field configuration
onto the nearest $Z(3)$ axis, with care
to preserve the sign. 
There are three distinct phases: a deconfined phase where
the projected expectation value satisfies $\left\langle Tr_{F}P\right\rangle >0$,
 a confined phase where $\left\langle Tr_{F}P\right\rangle =0$, 
and an intermediate phase with $\left\langle Tr_{F}P\right\rangle <0$,
which we refer to as the skewed phase. The locations of the phase
transitions were determined from the peaks of the adjoint Polyakov
loop susceptibility, and checked against the histograms of the fundamental
Polyakov loop. The dashed line in the phase diagram is an extrapolation;
the phase transition between the skewed and confined phases is very
difficult to resolve in this region. We will use the notation $H_{c1}$ for the
values of $H_A$ on the boundary between the deconfined and skewed phases,
and $H_{c2}$ for the boundary between the skewed and confined phases.

\begin{figure}
\centering
\includegraphics[width=.35\textwidth]{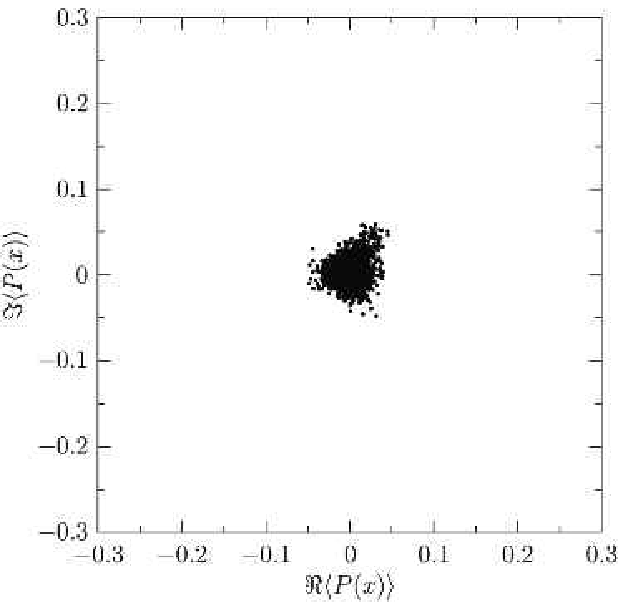}
\vspace{-5mm}
\caption{$SU(3)$ Polyakov loop histogram at $\beta=6.5$, $H_A=-0.1$.}
\label{fig:5}
%\end{figure}
\vspace{2mm}
%\begin{figure}
\centering
\includegraphics[width=.35\textwidth]{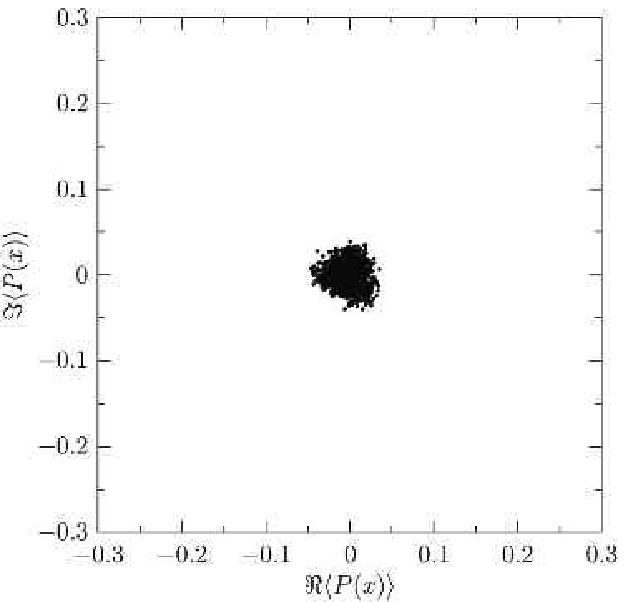}
\caption{$SU(3)$ Polyakov loop histogram at $\beta=6.5$, $H_A=-0.11$.}
\vspace{-5mm}
\label{fig:6}
\end{figure}

Figures \ref{fig:2} though \ref{fig:6} show 
histograms of the order parameter $\left\langle Tr_{F}P\right\rangle $
for various values of $H_{A}$ at  $\beta=6.5$. At $H_{A}=-0.06$,
there is clear evidence for the new intermediate phase where
 $\left\langle Tr_{F}P\right\rangle <0$. 
The skewed phase breaks $Z(3)$ symmetry, as shown clearly by the
histogram at $H_{A}=-0.08$, where all three possible skewed phases
appear. The appearance of significant tunneling between the three
phases on a $24^{3}\times4$ lattice is an indication that the transition
from the skewed phase to the confined phase is very weak. The skewed
phase differs from the deconfined phase not only in the orientation
of the histograms, but also in the smaller magnitude of $\left\langle Tr_{F}P\right\rangle $
for the skewed phase.
Near $H_{c1}$,
the orientation of fluctuations in histograms
of the skewed phase
is predominantly tangential,
but becomes more radial as $H_{c2}$
 is approached.
The transition between the deconfined and skewed phase is
clearly first-order, because the order parameter shows a marked jump
when changing sign. The transition between the skewed phase and the
confined phase is likely to be first order, because it is associated
with the universality class of the three-dimensional Potts model and
its generalizations via Svetitsky-Yaffe universality. However, in
simulations $\left\langle Tr_{F}P\right\rangle $ shows a very small
change at the skewed-confined transition, particularly near the apparent
tricritical point. Empirically, for a given value of $N_{t},$ the
skewed phase shows up clearly only for $N_{s}/N_{t}\geq6$. On a $12^{3}\times6$
lattice, for example, the skewed phase always appears to coexist
 with either the deconfined phase or the confined phase. A detailed
finite-size scaling analysis on very large lattices would be required
to resolve the order of this transition with confidence. 

\begin{figure}
\centering
\includegraphics[width=.55\textwidth]{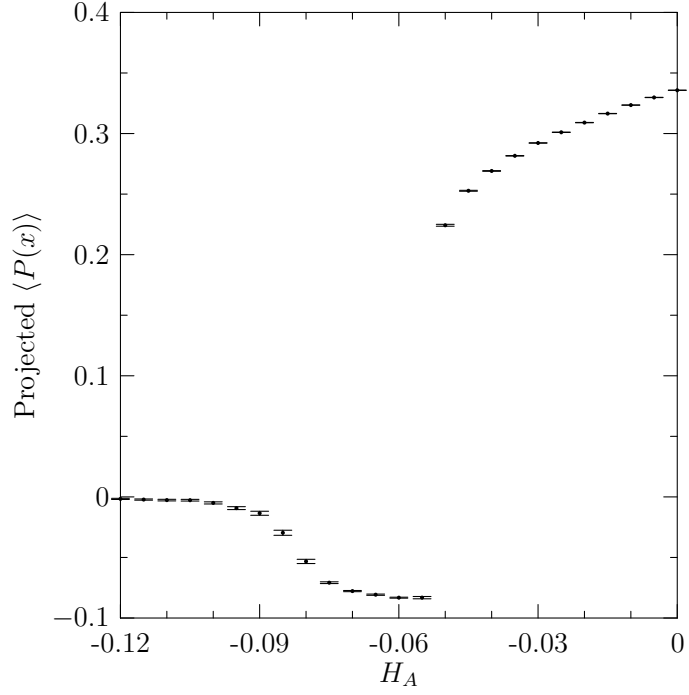}
\caption{Projected $Tr_{F}P$ for $\beta=6.5$.}
\label{fig:7}
\end{figure}

\begin{figure}
\centering
\includegraphics[width=.55\textwidth]{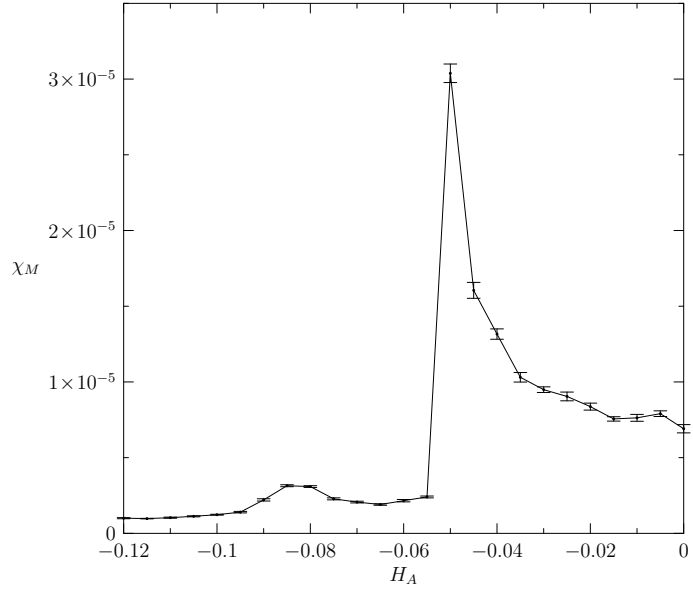}
\caption{Adjoint susceptibility $\chi_{M}$ for $\beta=6.5$.}
\label{fig:8}
\end{figure}

Figure
\ref{fig:7} shows the projected value of $\left\langle Tr_{F}P\right\rangle $
for various values of $H_{A}$ at $\beta=6.5$.
The presence of three distinct phases is clear.
The adjoint susceptibility $\chi_{M}$ for $\beta=6.5$ is shown in Fig. \ref{fig:8}.
There is a clear peak between the deconfined and skewed phases, and
a much smaller peak separating the skewed and confined phases.

\section{Theory for $SU(3)$}

A simple theoretical approach based on the effective potential $V_{eff}$
for Polyakov loop eigenvalues reproduces the phase structure observed
in simulations for $SU(3)$ and $SU(4)$. The effective potential
has two parts. The first part is the one-loop expression for the free
energy of gluons moving in a non-trivial, constant Polyakov loop background.
The one-loop free energy density was first evaluated by Gross, Pisarski,
and Yaffe \cite{Gross:1980br}, and by N. Weiss \cite{Weiss:1980rj}.
It is convenient to work in a gauge
where $A_{0}$ is a constant element of the $SU(N)$ Lie algebra so
that the background Polyakov loop is given simply by $P=\exp(i\beta A_{0})$.
The second contribution to the effective potential in our model is
simply the term $-h_{A}T\, Tr_{A}P$ that we have added to the gauge
Lagrangian. At temperature $T$, our expression for 
$V_{eff}$ is given by
\begin{equation}
V_{eff}=-2\frac{1}{2}Tr_{A}\int\frac{d^{3}k}{(2\pi)^{3}}T\sum_{n}\ln[(\omega_{n}-A_{0})^{2}+k^{2}]-h_{A}T\, Tr_{A}P\end{equation}
where the sum is over Matsubara frequencies $\omega_{n}=2\pi nT$.
A useful form is
%\begin{eqnarray}
%V_{eff}=\sum_{j,k=1}^{N}\left(1-\frac{\delta_{jk}}{N}\right)
%\left[-\frac{\pi^{2}T^4}{45}+\frac{T^4}{24\pi^{2}}\left|\Delta\theta_{jk}\right|^{2}\left(2\pi-\left|\Delta\theta_{jk}\right|\right)^{2}\right] \nonumber \\
%-h_{A}T\left(\left|\sum_{j=1}^{N}e^{i\theta_{j}}\right|^{2}-1\right)\end{eqnarray}
\begin{eqnarray}
\begin{aligned}
V_{eff}=\sum_{j,k=1}^{N}\left(1-\frac{\delta_{jk}}{N}\right)
\left[-\frac{\pi^{2}T^4}{45}+\frac{T^4}{24\pi^{2}}\left|\Delta\theta_{jk}\right|^{2}\left(2\pi-\left|\Delta\theta_{jk}\right|\right)^{2}\right]
& -h_{A}T\left(\left|\sum_{j=1}^{N}e^{i\theta_{j}}\right|^{2}-1\right)
\end{aligned}
\end{eqnarray}

 where the angles $\theta_{j}$ are the eigenvalues of $\beta A_{0}$
and$\left|\Delta\theta_{jk}\right|$ is $\left|\theta_{j}-\theta_{k}\right|mod\,2\pi.$
Thus $V_{eff}$ is the sum of a one-loop term plus another term treated
classically. 

The phase diagram is found by minimizing $V_{eff}$ as a function
of the Polyakov loop eigenvalues. The two terms that make up $V_{eff}$
have identical local extrema, and the problem of minimizing $V_{eff}$ can
be reduced to finding the minimum over this set. In the case of $SU(3)$,
it is sufficient to consider
$V_{eff}$ as $Tr_{F}P$
varies along the real axis. In this case, the eigenvalues of $P$
may be taken to be the set $\{1,\exp(i\phi),\exp(-i\phi)\}$,
and $Tr_{F}P$ may be written as $1+2\cos(\phi)$. The effective potential
is given by\begin{equation}
V_{eff}(\phi,T,h_A)=\frac{T^{4}}{6\pi^{2}}(8\phi^{2}(\phi-\pi)^{2}+\phi^{2}(\phi-2\pi)^{2})-h_{A}T((1+2\cos(\phi))^{2}-1).\end{equation}
The extrema of $V_{eff}$ occur at $\phi=0$, $\phi=2\pi/3$,
and $\phi=\pi$. The values of $Tr_{F}P$ for these values of $\phi$
are $3$, $0$, and $-1$, and we identify them with the deconfined,
confined, and skewed phases, respectively. The set of eigenvalues
$\{1,\exp(2\pi i/3),\exp(-2\pi i/3)\}$ is the unique set invariant
under global $Z(3)$ transformations \cite{Meisinger:2001cq,Schaden:2004ah}.

\begin{figure}
\centering
\includegraphics[width=.45\textwidth]{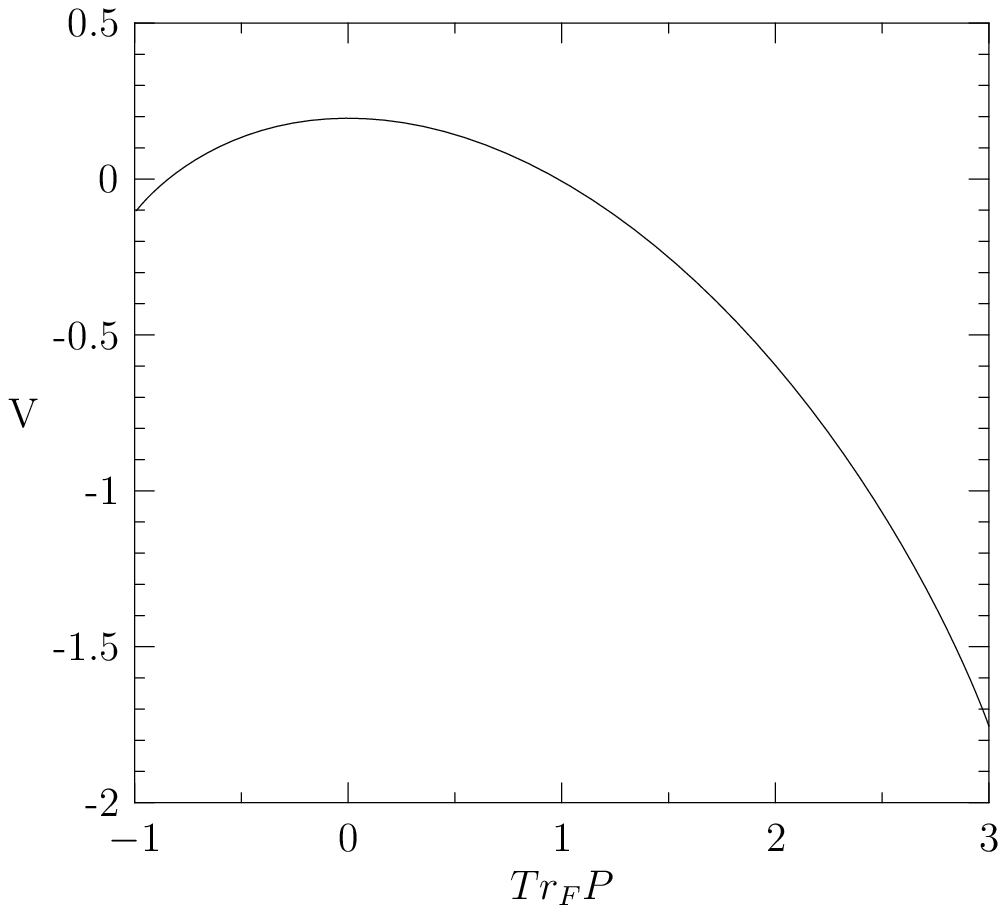}
\vspace{-5mm}
\caption{Effective potential versus $Tr_F P$ for deconfined phase at $h_A/T^3 =0$.}
\label{fig:9}
%\end{figure}
\vspace{2mm}
%\begin{figure}
\centering
\includegraphics[width=.45\textwidth]{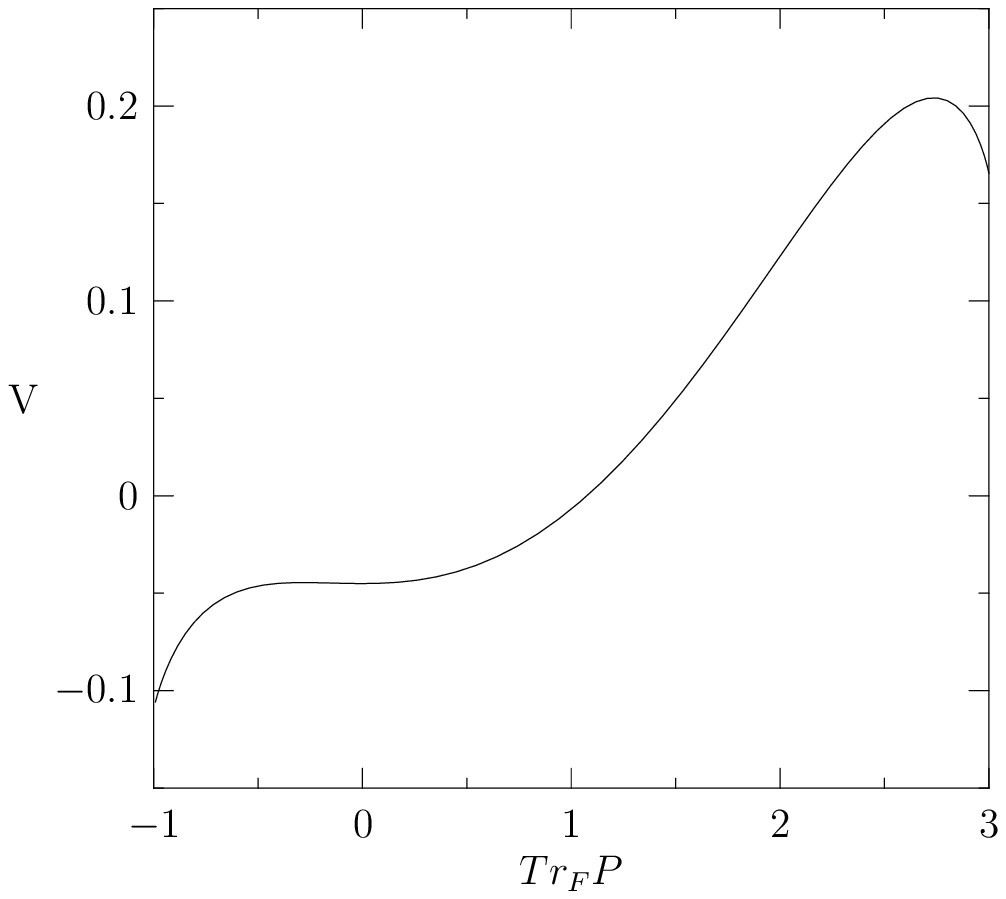}
\vspace{-5mm}
\caption{Effective potential versus $Tr_F P$ for skewed phase at $h_A/T^3 =-0.24$.}
\label{fig:10}
%\end{figure}
\vspace{2mm}
%\begin{figure}
\centering
\includegraphics[width=.45\textwidth]{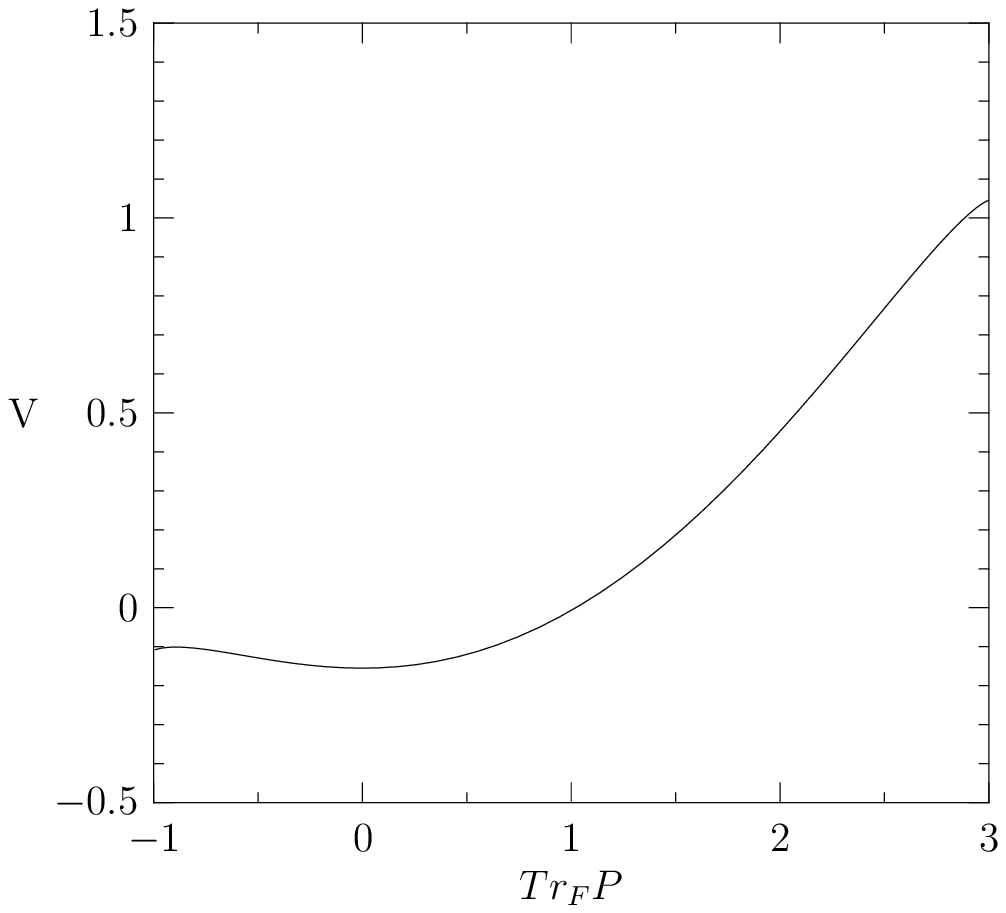}
\vspace{-5mm}
\caption{Effective potential versus $Tr_F P$ for confined phase at $h_A/T^3 =-0.35$.}
\label{fig:11}
\end{figure}

It is clear that the phase structure depends only on the dimensionless
variable $h_{A}/T^{3}$. As $h_{A}$ is lowered from zero, there is
a first-order transition from the deconfined phase to the skewed phase.
Setting the effective potential at $\phi=0$ and $\phi=\pi$ equal,
we find that the transition from the deconfined phase to the skewed
phase takes place at $h_{c1}/T^{3}=-\pi^{2}/48\simeq-0.206$. As $h_{A}$ decreases,
another first-order transition, this time between the skewed and confined
phases, occurs at $h_{c2}/T^{3}=-5\pi^{2}/162\simeq-0.305$ .We plot
the potential as a function of $Tr_{F}P$ for values in the three
phase in Figures \ref{fig:9}-\ref{fig:11}, corresponding to $H_{A}/T^3=0,-0.24,-0.35$.

We cannot directly relate $h_{A}$ and the corresponding lattice parameter
$H_{A}$, because there is an unknown multiplicative renormalization
relating the two. However, the ratio $h_{c2}/h_{c1}$ is approximately
$1.48$. If we assume that the relation of $h$ to $H$ is approximately
independent of $h$, we can compare with the results obtained from
simulation. 
As shown in Fig. \ref{fig:1}, the ratios $H_{c2}/H_{c1}$
obtained vary from $1.27$ at $\beta=6.2$ 
to $1.44$ at $\beta=6.8$, with a maximum value of $1.73$ in between. 

As noted previously, our simulations show a pronounced asymmetry in the skewed
phase between the fluctuations of the imaginary and the real parts
of $Tr_{F}P$. Fluctuations in the projected imaginary part are associated
with motion in the $\lambda_{8}$ direction, while fluctuations in
the projected real part are due to motion in both the $\lambda_{8}$
and $\lambda_{3}$ directions. It is thus interesting that in the
skewed phase, theory predicts an asymmetry in the screening masses
obtained from small fluctuations in the eigenvalues of $P$. This
is quite different from the behavior in the confined and deconfined
phases, where theory predicts no asymmetry. We have
\begin{equation}
\frac{m_{3}}{m_{8}}=\sqrt{\frac{1+2h_{A}/T^{3}}{-1-6h_{A}/T^{3}}}
\end{equation}
This ratio varies from $1.59$ at $h_{c1}$ to $0.69$ at $h_{c2}$.
This is on the order of the variation seen in the fluctuations of
the real and imaginary parts of $Tr_{F}P$, and probably accounts
for the behavior seen in the histograms. This prediction for the mass
ratio can be checked more directly by comparing the masses obtained
from the correlation functions of the real and imaginary parts of
the projected Polyakov loop in the skewed phase.

\begin{figure}
\centering
\includegraphics[width=.55\textwidth]{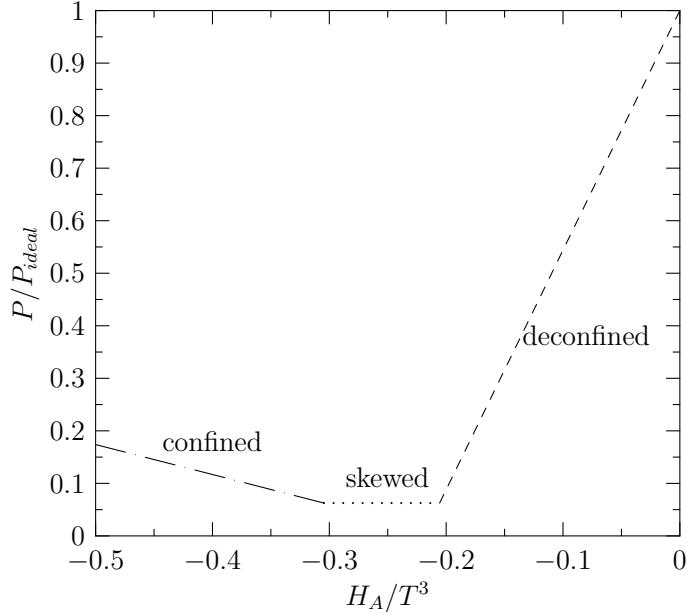}
\caption{Theoretical prediction for pressure normalized to black body
pressure pressure as a function of $h_A$.
}
\label{fig:12}
\end{figure}

The pressure can be calculated from simulations along a path of constant
$\beta$, using\begin{equation}
\frac{p_{2}}{T^{4}}-\frac{p_{1}}{T^{4}}=N_{t}^{3}\int_{1}^{2}dH_{A}\langle Tr_{A}P\rangle\end{equation}
A detailed comparison of the pressure for all values of $h_{A}$ would
require knowledge of the relation between $h_{A}$ and $H_{A}$. However,
it is relatively simple to compare the change in the pressure from
$h_{A}=H_{A}=0$ to the deconfined-skewed phase boundary as well as
the change in pressure across the skewed phase. 
Using $V_{eff}$, we find that the predicted
change in $p/T^{4}$ from $h_{c1}$ to $0$ is $\pi^{2}/6\simeq1.64$;
from $h_{c2}$ to $h_{c1}$ the net change is $0$. For comparison,
the corresponding results from simulations at $\beta=6.5$ are $1.64\pm0.03$
and P = $-0.18 \pm 0.07$. In each case, the error is completely
dominated by systematic error due to uncertainty in the location
of the critical values of $H_{A}$, with statisitical error at least an order
of magnitude smaller.

\section{Simulation results for $SU(4)$}

We have also simulated $SU(4)$ lattice gauge theories, again primarily
on $24^{3}\times4$ lattices. As in the case of $SU(3)$, we find
a new phase in the region $h_{A}<0$, but the nature of the new phase
is completely different. In this new, partially confined phase, global
$Z(4)$ symmetry is spontaneously broken to $Z(2)$. In this phase,
particles in the fundamental representation ({}``$SU(4)$ quarks'')
are still confined, but bound states of two such particles ({}``$SU(4)$
diquarks'') are not. Each irreducible representation of $SU(N)$
has an $N$-ality: if $z\in Z(N)$, $P\rightarrow zP$ induces a
change $Tr_{R}P\rightarrow z^{k}Tr_{R}P$, where $k$ is the $N$-ality
of the representation $R$. The characteristic feature of the partially
confined phase in $SU(4)$ is that the expected value of Polyakov
loops in $k=1$ representations is zero, but not in $k=2$ representations
such as as the ${\bf 6}$ and the ${\bf 10}$.

\begin{figure}
\centering
\includegraphics[width=.35\textwidth]{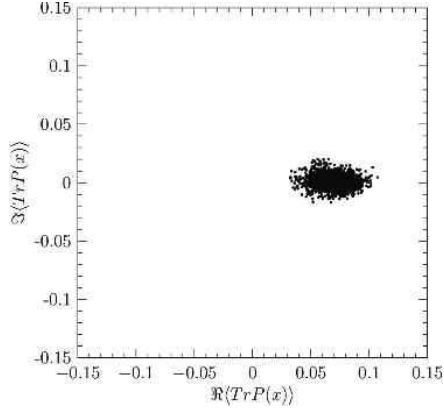}
\vspace{-5mm}
\caption{$SU(4)$ Polyakov loop histogram at $\beta=11.1$, $H_A=-0.1$.}
\label{fig:13}
\end{figure}

\begin{figure}
\centering
\includegraphics[width=.35\textwidth]{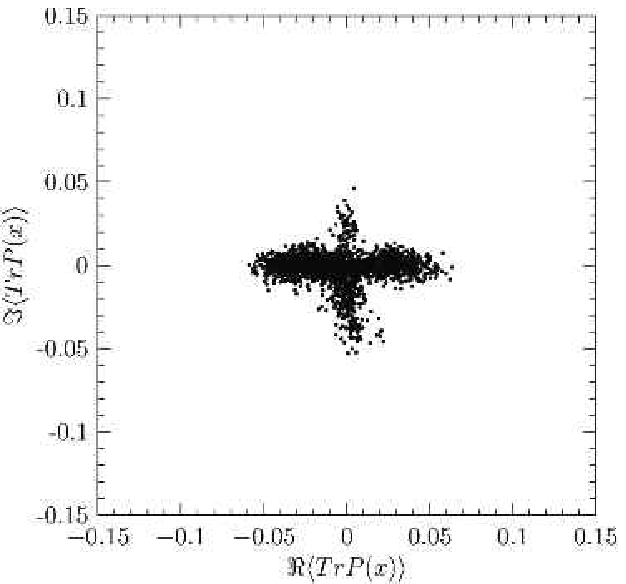}
\vspace{-5mm}
\caption{$SU(4)$ Polyakov loop histogram at $\beta=11.1$, $H_A=-0.11$.}
\label{fig:14}
\end{figure}

\begin{figure}
\centering
\includegraphics[width=.35\textwidth]{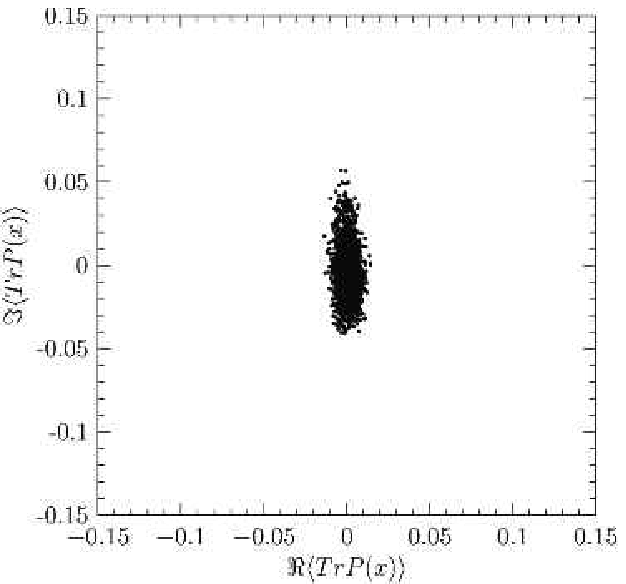}
\vspace{-5mm}
\caption{$SU(4)$ Polyakov loop histogram at $\beta=11.1$, $H_A=-0.12$.}
\label{fig:15}
\end{figure}

\begin{figure}
\centering
\includegraphics[width=.35\textwidth]{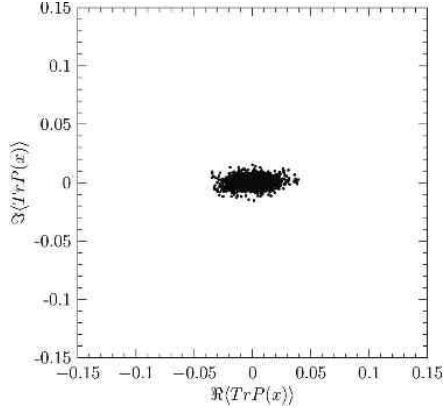}
\vspace{-5mm}
\caption{$SU(4)$ Polyakov loop histogram at $\beta=11.1$, $H_A=-0.125$.}
\label{fig:16}
\end{figure}

\begin{figure}
\centering
\includegraphics[width=.8\textwidth]{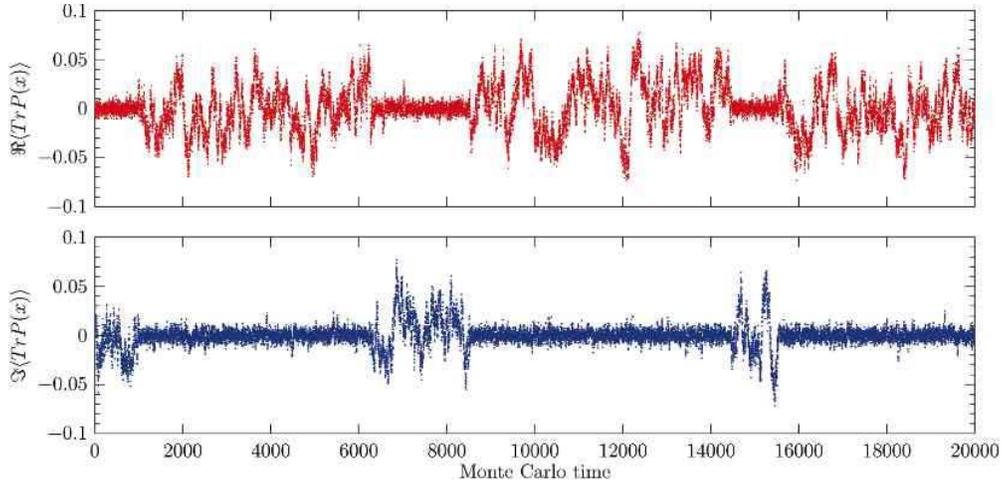}
\caption{Real and imaginary parts of $SU(4)$ Polyakov loop versus Monte Carlo time
at $\beta=11.1$, $H_A=-0.11$.}
\label{fig:17}
\end{figure}

The breaking of $Z(4)$ down to $Z(2)$ for sufficiently negative $H_A$
 is manifest in histograms of
the Polyakov loop in the fundamental representation as a clustering
of data around either the $x$ or $y$ axis, but not both, as shown
in Figures \ref{fig:13}-\ref{fig:16}. The $Z(2)$ character of this new phase is very
clearly shown in Figure \ref{fig:17}, which shows the behavior of the real
and imaginary part of the Polyakov loop versus Monte Carlo time for
one long run with 20,000 measurements. As the figure reveals, there
are significant fluctuations in either the real or the imaginary part,
but not both simultaneously, characterisitic of $Z(4)$ breaking to
$Z(2)$. In this phase, the expectation value of $Tr_{F}P^{2}$ is non-zero,
being positive when the fluctuations in $Tr_{F}P$ are along the real
axis, and negative when $Tr_{F}P$ fluctuates along the imaginary axis.

As $H_{A}$ becomes more negative. the histograms show decreasing
amplitude in the fluctuations of $Tr_{F}P$. It is possible that there
is a second phase transition from the $Z(2)$ phase to the confined
phase as $H_{A}$ becomes more negative, but we have not found direct
evidence for this. As we discuss below, our simple theoretical model does not
predict a second transition for this theory, at least not  at high temperatures where
it is valid.

\section{Theory for $SU(4)$}

We have examined within our simple theoretical model
 the possible occurrence of four different phases
in $SU(4)$: the confined phase, which has full $Z(4)$ symmetry;
the deconfined phase; a partially-confined, $Z(2)$-invariant phase;
and a skewed phase similar to the skewed phase of $SU(3)$. Only the
deconfined phase and the $Z(2)$ phase are predicted by our simple
theoretical model. 

The properties of the $Z(2)$-invariant phase may be understood by
considering the one-parameter class of eigenvalues invariant under
$Z(2)$; the eigenvalues in this class may be written as $\left\{ \theta,\pi-\theta,\pi+\theta,2\pi-\theta\right\} $,
and the corresponding Polyakov loops have the form $diag\left[e^{i\theta},-e^{-i\theta},-e^{i\theta},e^{-i\theta}\right]$.
The one-loop effective potential 
as a function of $\theta$ becomes 
\begin{equation}
V_{eff}=-\frac{\pi^{2}T^{4}}{3}+\frac{T^{4}}{6\pi^{2}}\left[\left(\pi-2\theta\right)^{2}\left(\pi+2\theta\right)^{2}+\pi^{4}+\left(2\pi-2\theta\right)^{2}\left(2\theta\right)^{2}\right]+h_{A}T\end{equation}
which has its minimum within this class at $\theta=0$. The confined
phase, which has $Z(4)$ symmetry, is realized at $\theta=\pi/4$
but is never the minimum of $V_{eff}$. This behavior is easy to understand:
both the confined and $Z(2)$-invariant phases have the same dependence
on $h_{A}$, so the stable phase is the one that minimizes the contribution
of the gauge bosons. The deconfined phase does not fall into the $Z(2)$-invariant
class: with all eigenvalues set to $0$, the value of the effective
potential in the deconfined phase is\begin{equation}
V_{d}=-\frac{\pi^{2}T^{4}}{3}-15h_{A}T\end{equation}
 There is a first-order transtion between the deconfined and $Z(2)$-invariant
phases at $h_{A}/T^{3}=-\pi^{2}/48\simeq-0.205617$. The value of
$\Delta\left(p/T^{4}\right)$ between $h_{A}=0$ and the critical
point is $\pi^{2}/3\simeq3.289$. The value we
obtained from simulations at $\beta=11.0$ was $2.21\pm0.07$,
where again the systematic error dominates, due to uncertainty
in the location of the transition. 

In order to realize
the confined phase, it may be necessary
 to add an additional term proportional to 
$Tr_{A}P^{2}=Tr_{F}P^{2}Tr_{F}P^{+2}-1$
in order to force
both $Tr_{F}P$ and $Tr_{F}P^{2}$ to zero,
but this has not yet been checked in
simulations. 
Of course,
there must be a line of transitions in the
$\beta-H_{A}$ plane separating the $Z(2)$ phase from the low-temperature
confined phase with $Z(4)$ symmetry. The transition could be either
first or second order. We have not yet mapped out this phase boundary
via simulation. As previously noted, our simple theoretical model does not
include a mechanism for this transition.

\section{Conclusions}

We have considerable evidence, from lattice simulation and from theory,
for the existence of new phases of finite temperature gauge theories,
and for the restoration of the confined phase at high temperatures
when extra, $Z(N)$-invariant, Polyakov loop terms are added to the gauge action. In $SU(3)$,
a novel skewed phase was found, and in $SU(4)$, we found a phase
where $Z(4)$ is spontaneously broken to $Z(2)$. In the general case
of $SU(N)$, there is good reason to expect a very rich phase structure
 may exist.

A simple theoretical model based on perturbation theory at high temperatures
has proven surprisingly accurate in predicting the observed phase
structure and thermodynamics. Although successful, the model has significant
shortcomings. Fluctuations in $A_{0}$ are not considered, nor is
the renormalization of $h_{A}$.
Most importantly, our simple model
does not include  in $V_{eff}$ whatever mechanism
is responsible for confinement
at low temperatures when $h_{A}=0$.
It is therefore invalid at low temperatures.
Nevertheless, theory and simulation are in reasonable agreement on
a wide range of properties. Our model can also make predictions for
string tensions and 't Hooft loop surface tensions, and these predictions
can be checked in lattice simulations.
The introduction of $h_A$ as an extra parameter also
affects the action of calorons, topologically stable
solutions of the classical field equations, and thus
may offer rich possibilities for explorations in instanton
physics. 

The interpretation of these additional phases of finite temperature
gauge theories is, to a degree, associated with the issue of a physical
implementation of a negative value for $h_{A}$. The existence of
a partially confining, $Z(2)$-invariant phase in $SU(4)$ might have
been expected \cite{Meisinger:2001cq},
 and the interpretation of the order parameters is clear.
The interpretation of the skewed phase in $SU(3)$ is less certain.
As in the deconfined phase, the global symmetry of the Polyakov loop
is lost in the skewed phase. Our theoretical analysis indicates that,
on average, two of the three Polyakov loop eigenvalues are degenerate,
suggesting a possible interpretation of the skewed
phase as some form of $SU(2)\times U(1)$ Higgs
phase.

The issues underlying the interpretation of parameters and phases is connected with
 the association of finite temperature gauge theories with universality
classes of spin systems \cite{Svetitsky:1982gs}. It has always been assumed implicitly
that the mapping from gauge theories to spin systems 
is into but perhaps not onto. There are phases of
$SU(N)$ and $Z(N)$ spin systems which are not easily obtainable
from physical finite temperature gauge theories. For example, the
antiferromagnetic phase of a spin system can be obtained from the
strong-coupling effective action of a lattice gauge theory with $g^{2}<0$
and $N_{t}$ odd, a construction with no obvious continuum
limit. However, phases can often be reached in different ways in
the space of parameters.
The skewed phase we have found in $SU(3)$ gauge theory is
very similar to the anti-center phase found in $SU(3)$ spin systems
by Wozar {\it et al.} \cite{Wozar:2006fi}. Although the term in the spin Hamiltonian that
produces the anti-center phase is associated with the ${\bf 15}$ representation
rather than the adjoint term we have used, we are confident that the
two phases will prove to be related. At this time, it is simply unclear what physical
principles, if any, limit the map between spin systems and gauge theories.

We believe that the ability to create new phases in a controlled way
may become an important tool in understanding the properties of finite
temperature gauge theories. For example, simulations indicate that
the confined phase obtained in $SU(3)$ at high $T$ with $h_{A}<0$
is connected to the conventional confined phase at low $T$ with $h_{A}=0$.
The possibility of a confined phase in a region where perturbation
theory is valid is by itself enormously interesting. As larger gauge
groups are considered, the number of possible new phases increases.
For example, in $SU(6)$, we can consider partial breaking of $Z(6)$
to either $Z(2)$ or $Z(3).$ At high temperatures, analytic calculations
of both string tensions and 't Hooft loop surface tensions can be
carried out in these different phases for potential comparison
with simulation \cite{MeMyMCO:2007re}.

\bibliography{hA_working}
%\begin{thebibliography}{1}
%\bibitem{fj}de Forcrand, P. and O. Jahn (2005). Monte Carlo Overrelaxation
%for SU(N) Gauge Theories; arXiv: hep-lat/0503041.
%\end{thebibliography}

\end{document}